\documentclass[twocolumn,aps,prb,showpacs,superscriptaddress]{revtex4}

\usepackage{amsmath,bm}
\usepackage{graphicx,epsfig,amssymb,amsfonts}

\newcommand{\ba}{\begin{eqnarray}}
\newcommand{\ea}{\end{eqnarray}}
\newcommand{\be}{\begin{equation}}

\newcommand{\ee}{\end{equation}}
\newcommand{\bd}{\begin{displaymath}}
\newcommand{\ed}{\end{displaymath}}
\renewcommand{\v}[1]{{\bf #1}}

\newcommand{\ra}{\rangle}
\newcommand{\la}{\langle}

\newcommand{\bpm}{\begin{pmatrix}}
\newcommand{\epm}{\end{pmatrix}}

\newcommand{\fbar}{f^+}
\newcommand{\chibar}{\chi^*}

\begin{document}

\title{Chiral spin states in the pyrochlore Heisenberg magnet}

\author{Jung Hoon Kim}
\affiliation{Department of Physics, BK21 Physics Research Division,
Sungkyunkwan University, Suwon 440-746, Korea}
\author{Jung Hoon Han}
\email[Electronic address:$~$]{hanjh@skku.edu}
\affiliation{Department of Physics, BK21 Physics Research Division,
Sungkyunkwan University, Suwon 440-746, Korea}

\begin{abstract} Fermionic mean-field theory and variational Monte
Carlo calculations are employed to shed light on the possible uniform
ground states of the Heisenberg model on the pyrochlore lattice.
Among the various flux configurations, we find the chiral spin states
carrying $\pm \pi/2$ flux through each triangular face to be the most
stable, both within the mean-field theory and the projected wave
function studies. Properties of the spin-spin correlation function
and the chirality order parameter are calculated for the projected
wave functions. Mean-field band structures are examined.
\end{abstract}


\maketitle

\textbf{Introduction}: The question of the quantum ground state of
spin-$1/2$ Heisenberg Hamiltonian
\ba H = \sum_{\la ij \ra} \v S_{i}\cdot \v S_{j}
\label{eq:Heisenberg-H}\ea
on the pyrochlore lattice has prompted active research for nearly two
decades\cite{lacroix,harris,tsunetsugu,berg,moessner,shankar}. A
popular scheme, employed in Refs.
\onlinecite{lacroix,harris,tsunetsugu,berg}, was to solve the
isolated tetrahedron problem exactly and to couple the nearby
disjoint tetrahedra in the weak  exchange energy $J'$. A non-magnetic
state with exponentially short correlation length was identified in
Ref. \onlinecite{lacroix}, and Refs.
\onlinecite{harris,tsunetsugu,berg} noted the dimer instability in
the ground state. A similar dimer instability was discovered in the
large-$N$ approach in Ref. \onlinecite{moessner}. While Refs.
\onlinecite{harris,tsunetsugu,berg} begin with the dimer basis of a
single tetrahedron solution to carry out perturbation in $J'$, Ref.
\onlinecite{moessner} starts with a translationally invariant
solution and finds that dimerization occurs as a spontaneous symmetry
breaking.

In this paper, we adopt the fermionic mean-field followed by
variational Monte Carlo (VMC) treatment to address this issue. Prior
work in the same spirit for the Kagome lattice can be found in
Ref.~\onlinecite{ran}. We find chiral spin states\cite{wwz} with
non-zero averages of the three-spin operator $\langle \v S_i \cdot \v
S_j \times \v S_k \rangle$ for the elementary triangular unit formed
by $\langle ijk\rangle$ sites to be the likely non-magnetic ground
state of the Heisenberg model realized on pyrochlore lattice, under
the assumption of uniform nearest-neighbor bond amplitudes $| \langle
\v S_i \cdot \v S_j \rangle |$. The flux through the triangles are
found to be $\pm \pi/2$ at the mean-field level, but reached a
smaller value after the Gutzwiller projection was carried out.

\begin{widetext}
\begin{figure*}[t]
\includegraphics[scale=0.8]{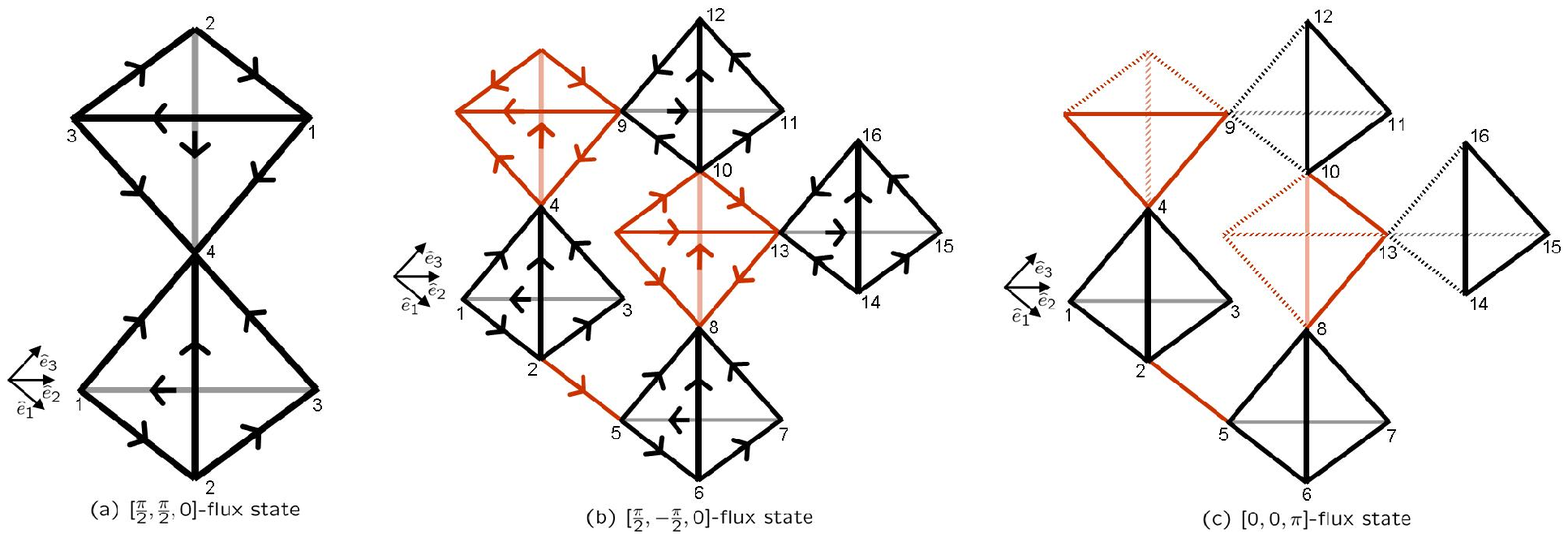}
\caption{ (color online) $\chi_{ij}$ bond configurations  producing
(a) $[\frac{\pi}{2},\frac{\pi}{2},0]$, (b)
$[\frac{\pi}{2},-\frac{\pi}{2},0]$ and (c) $[0,0,\pi]$ flux states.
Setting $| \chi_{ij} | =1$, the direction of the arrows indicates
$\chi_{ij} = +i$ going from $i$ to $j$.  The shaded bonds are lying
on the back sides. Independent sites are labeled $1$ through $4$
(one up tetrahedron in a unit cell) in (a) and 1 through 16 (four up
tetrahedra in a unit cell) in (b) and (c). In (c), the thick(dotted)
lines represent bonds with $\chi_{ij} = 1 (-1)$.}
\label{fig:flux-configs}
\end{figure*}
\end{widetext}

\textbf{Mean-field Theory}: In rewriting the spin operator as a
fermion bilinear and introducing the mean-field variable $\chi_{ij}
= \langle \fbar_{i} f_{j} \rangle$, one arrives at the mean-field
Hamiltonian

\be H_\mathrm{MF} = -\sum_{i} \sum_{j\in i} \chibar_{ij} \fbar_{i}
f_{j} + \sum_{i} \lambda_i (\fbar_{i} f_{i}
-1/2),\label{eq:mean-field-H}\ee
with the Lagrange multiplier $\lambda_i$ enforcing the occupation
constraint at each site. In doing the mean-field calculation we drop
the spin index and work with the half-filled case $\langle \fbar_i
f_i \rangle = 1/2$. The summation over all nearest-neighbor sites $j$
with respect to $i$ is indicated by $\sum_{j \in i}$.  To set up the
coordinates, we place the four corners of a single ``up" tetrahedron
(uT) at $(0,0,0)$, $(1,0,0)$, $(1/2,\sqrt{3}/2,0)$,
$(1/2,1/2\sqrt{3},\sqrt{2/3})$, then displace it by integer
combinations of $\hat{e}_1 = (2,0,0)$, $\hat{e}_2 = (1,\sqrt{3},0)$,
and $\hat{e}_3 = (1,1/\sqrt{3},\sqrt{8/3})$ to generate the
pyrochlore lattice: $n_1 \hat{e}_1 + n_2 \hat{e}_2 + n_3 \hat{e}_3$,
$n_i$=integers. Each uT is met at four corners by down tetrahedra
(dT) and vice versa as shown in Fig. \ref{fig:flux-configs}. The
self-consistent mean-field calculations were run for $L\times L\times
L \equiv L^3$ lattice with $4L^3$ lattice sites. $L$ refers to the
number of uT's along each $\hat{e}_\alpha$ direction. The lattice
contains an equal number of uT and dT blocks.

A completely unrestricted minimization of $\langle
H_\mathrm{MF}\rangle$ resulted in the ground state with each site
paired into a dimer while all dimers are disconnected from one
another, in accordance with Rokhsar's general
observation\cite{rokhsar}. The extensive degeneracy of the dimer
ground state will be lifted at higher orders in $1/N$ in a large-$N$
expansion to give rise to a ground state with (possibly) restored
translational symmetry. Furthermore, the fully dimerized state
carries an energy of $-0.375$ per site, much higher than some of the
uniform states we consider in this paper. Therefore, we move with the
idea that the true ground state of Eq. (\ref{eq:Heisenberg-H}) is
better captured by the uniform amplitude ansatz $|\chi_{ij}|=\chi$,
but with arbitrary phases: $\chi_{ij} = \chi e^{i\phi_{ij}}$. For a
single tetrahedron, such a search yielded solutions where the flux
through the four triangular faces are all equal to $\Phi= +\pi/2$ or
$\Phi= -\pi/2$. The flux is defined from the directed product
$e^{i\Phi}= e^{i[\phi_{ij}+\phi_{jk}+\phi_{ki}]}$ as the three sites
of a triangle $\langle ijk \rangle$ are traversed in a
counterclockwise manner as viewed from outside the tetrahedron. For
the lattice problem with uniform $|\chi_{ij}|$, the mean-field ground
state is found to be the one with staggered chirality: $\Phi=\pi/2$
for all the uT's and $\Phi=-\pi/2$ for all the dT's, or vice versa.
No solutions were found where different faces of a given tetrahedron,
either up or down, carried different amounts of flux, or flux other
than $\pm \pi/2$. The flux through the hexagons of the pyrochlore
lattice were all zero. The mean-field ground state solution is in
complete conformity with the ``Rokhsar rule"\cite{rokhsar-rule} of
the flux for different types of polygons.

\textbf{Variational Monte Carlo Calculation}: Variational Monte Carlo
(VMC) calculations of the energies of several mean-field ansatz
states were carried out including the one found in the mean-field
calculation. The states are labeled by $[uT,dT,H]$, where the three
numbers $uT$, $dT$, and $H$ refer to the flux through the triangles
of the up tetrahedra, down tetrahedra, and the hexagons,
respectively. We examine four such states, (i) $[0,0,0]$, (ii)
$[{\pi\over 2},{\pi\over 2},0]$, (iii) $[{\pi\over 2},-{\pi\over
2},0]$, (iv) $[0,0,\pi]$. The mean field result corresponds to case
(iii). For future reference, we denote (ii) and (iii) as uniform and
staggered flux states. The other two states, (i) and (iv), do not
break time reversal symmetry. The choice of the trial states are
motivated by and parallels those in Ref.~\onlinecite{ran} for the
Kagome lattice. The $\chi_{ij}$ bond patterns that generate each
state at the mean-field level are shown in
Fig.~\ref{fig:flux-configs}. With $\chi_{ij}$'s given in Fig.
\ref{fig:flux-configs} as input to Eq. (\ref{eq:mean-field-H}) one
can diagonalize $H_\mathrm{MF}$ to obtain the mean-field ground state
$|\psi_\mathrm{MF}\rangle$ as a Slater determinant. The evaluation of
the energy and other operators $\hat{X}$ are carried out in the
projected space $|s\rangle$ ($|s\rangle$ spans all the states with
one spin per site) by

\ba \langle \hat{X}\rangle = { \sum_s \langle \psi_\mathrm{MF} | s
\rangle \langle s | \hat{X}|\psi_\mathrm{MF}\rangle \over \sum_{s}
\langle \psi_\mathrm{MF}| s \rangle \langle s|
\psi_\mathrm{MF}\rangle }=\sum_s P(s) {\langle s | \hat{X}
|\psi_\mathrm{MF} \rangle \over \langle s | \psi_\mathrm{MF}\rangle
},\ea
where $P(s) = |\langle s | \psi_\mathrm{MF} |^2 /\sum_s |\langle s |
\psi_\mathrm{MF} \rangle |^2$ is the probability weight used in the
Monte Carlo procedure\cite{gros}.

The unit cell includes a single tetrahedron for the [0,0,0] and
[${\pi\over 2},{\pi\over 2},0$] states, and $2\times 2\times 1$
tetrahedra for the $[{\pi\over 2},-{\pi\over 2},0]$ and $[0,0,\pi]$
states. Periodic boundary conditions (PBC) generate degenerate states
at the Fermi level for the $[0,0,0]$ and $[{\pi\over 2},{\pi\over
2},0]$ cases, which can be lifted by applying anti-periodic boundary
conditions (aPBC) along one of the directions, \textit{e.g.}
$\hat{e}_1$, for $[0,0,0]$, and along two directions, \textit{e.g.}
$\hat{e}_1$ and $\hat{e}_2$, for $[{\pi\over 2},{\pi\over 2},0]$. In
the case of $[{\pi\over 2},-{\pi\over 2},0]$ and $[0,0,\pi]$ the unit
cell includes four up tetrahedra, two in the $\hat{e}_1$-direction
and two in the $\hat{e}_3$-direction, for the $\chi_{ij}$ patterns
shown in Fig.~\ref{fig:flux-configs}(b) and (c). PBC suffices in both
these cases since no degeneracy occurs at the Fermi level.

The VMC energies of the trial states we propose are listed in
Table~\ref{table:VMC-energy}. Several independent Monte Carlo
simulations were made with each simulation consisting of over $10^5$
steps (each step means one sweep of the whole lattice size) for each
state to obtain reliable estimates of the energy. Due to the rapid
increase of the system size with $L$, the calculation is currently
limited to $L=4$. For $L$ up to $4$, the two chiral states turn out
to have much lower energies than the two non-chiral states. The
energies of the two chiral states are very close.

\begin{table}[ht]
\caption{Energy per site of various states obtained from VMC
depending on the number of up tetrahedra $L$ in each direction. The
two chiral states have much lower energies than the non-chiral
states. Statistical uncertainties lie below the digits shown.
Boundary conditions, aPBC or PBC, used in the calculation are
listed.} \centering
\begin{tabular}{c c c c c}
\hline $L$ & $[0,0,0]$ & $[{\pi\over 2}, {\pi\over 2},0]$ &
$[{\pi\over2},-{\pi\over 2},0]$ & $[0,0,\pi]$ \\
\hline
   & aPBC & aPBC & PBC & PBC \\
2  & -0.372 & -0.478 & -0.466 & -0.374 \\
4  & -0.374 & -0.459 & -0.456 & -0.375 \\
\hline
\end{tabular}
\label{table:VMC-energy}
\end{table}

\textbf{Spin-spin Correlation and Chirality}: The spin-spin
correlation function can be computed using the relation $\langle \v
S_i \cdot \v S_j \rangle = (3/4)\langle \sigma_i^z \sigma_j^z
\rangle$ due to the spin isotropy of the ground state. Table
\ref{table:spspcor} displays the spin-spin correlations obtained for
all four flux configurations as a function of the distance for $L=4$.
We measure the correlation along the six directions, $\hat{e}_1$,
$\hat{e}_2$, $\hat{e}_3$, $\hat{e}_1-\hat{e}_2$,
$\hat{e}_2-\hat{e}_3$, $\hat{e}_3-\hat{e}_1$, and take averages. The
results for any one direction is consistent with those taken along
any other. It was difficult to reach a large enough size to
discriminate an algebraic decay of the correlation against an
exponential one. The sign of $\langle \v S_i \cdot \v S_j \rangle$
alternated with distance as expected in an antiferromagnetically
correlated state, except for $[0,0,\pi]$ where the signs of the
second- and the third-neighbor correlations are reversed (See
Table~\ref{table:spspcor}). The fast decay of the spin-spin
correlation is consistent with an earlier finding of Ref.
\onlinecite{lacroix} based on the cluster perturbation method.

\begin{table}[ht]
\caption{The spin-spin correlation function $\la \v S_i \cdot \v S_j
\ra$ with respect to distance for the four flux states. Five
independent $10^5$ MC steps were used for each data point.}
\centering
\begin{tabular}{c c c c c}
\hline $|i\!-\!j|$ & $[0,0,0]$ & $[{\pi\over 2}, {\pi\over 2},0]$ &
$[{\pi\over2},-{\pi\over 2},0]$ & $[0,0,\pi]$ \\
\hline
1  & -0.1249(2) & -0.1533(1) & -0.1522(1) & -0.1250(1) \\
2  & +0.0088(3) & +0.0194(2) & +0.0122(1) & -0.0054(3) \\
3  & -0.0027(1) & -0.0129(1) & -0.0024(3) & +0.0012(2) \\
4  & +0.0007(2) & +0.0060(2) & +0.0006(3) & +0.0001(6) \\
\hline
\end{tabular}
\label{table:spspcor}
\end{table}

Both the mean-field search and the VMC calculation suggest that the
chiral states with spontaneously broken time-reversal symmetry
($\mathcal{T}$) is vital in the understanding of the ground state
correlation of the Heisenberg model on the pyrochlore lattice. We
develop below an extension of the VMC method which allows the
calculation of the chirality and evaluate it for the chiral states.

\begin{figure}[ht]
\begin{center}
\includegraphics[scale=0.4]{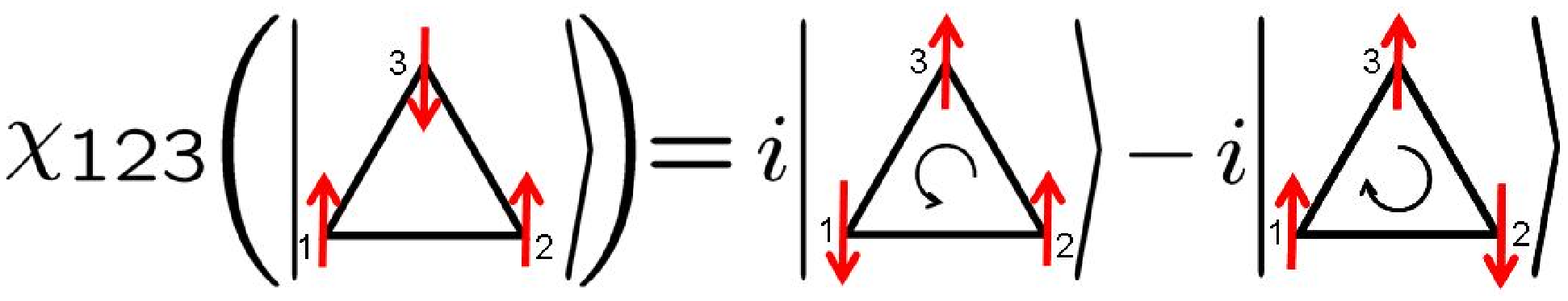}
\end{center}
\caption{(color online) The two states, $\left| s_+ \right.\ra$ and
$\left| s_- \right.\ra$, produced by the chirality operation on a
state $\left| s \right.\ra$.} \label{fig:chirality}
\end{figure}

The scalar chirality $\hat{\chi}_{123}=\langle \v S_1 \cdot \v S_2
\times \v S_3 \rangle$ is equivalent to\cite{wwz}

\be {1\over 2}\mathrm{Im} (\langle
\hat{\chi}_{12}\hat{\chi}_{23}\hat{\chi}_{31} \rangle ) = { \langle
\hat{\chi}_{12}\hat{\chi}_{23}\hat{\chi}_{31} -
\hat{\chi}_{13}\hat{\chi}_{32}\hat{\chi}_{21} \rangle \over 2i} . \ee
with $\hat{\chi}_{ij}=\sum_\sigma \fbar_{i\sigma}f_{j\sigma}$. It is
the difference of the cyclic permutation of the spins done in the
counterclockwise (CCW) and clockwise (CW) directions. Unlike in the
evaluation of the spin-spin correlation where the action of
$\sigma_i^z \sigma_j^z$ on a given basis state $|s\rangle$ is
diagonal, the outcome of $\hat{\chi}_{123}|s\rangle$ is not
proportional to $\left|s \right.\ra$ itself. Instead, we obtain the
relation $\hat{\chi}_{123}\left| s \right.\ra =i ( \left| s_{+}
\right.\ra - \left| s_{-} \right.\ra)$, as depicted in
Fig.~\ref{fig:chirality}. The two states $\left| s_{+} \right.\ra$
and $\left| s_{-} \right.\ra$ are obtained as CCW and CW rotations by
one lattice site of the original spin configuration $\left| s
\right.\ra$ with the rotation axis chosen to point out of the
triangular faces of the tetrahedra. When the state $|s\rangle$
contains all three spins up or all down for a given triangle, the
chirality operation gives zero. The average $\langle
\hat{\chi}_{123}\rangle $ is obtained from

\ba && \la \hat{\chi}_{123} \ra = i \times \sum_{s}P(s)\left[
\frac{\la s_-|\psi \ra - \la s_+|\psi \ra } { \la s|\psi \ra }
\right]. \label{eq:chirality-formula}\ea
Here $P(s)$ is the statistical weight $|\langle s |\psi\rangle |^2
/\sum_{s} |\langle s | \psi \rangle |^2$ for a given mean-field state
$|\psi\rangle$. Following the usual manner of updating the
configuration by Monte Carlo methods, one has to calculate the ratio
$(\la s_-|\psi \ra - \la s_+|\psi \ra ) / \la s|\psi \ra$ for each
state $\left| s \right.\ra$.

\begin{table}[ht]
\caption{The real and imaginary parts of the averages of $\la
s_+|\psi \ra / \la s|\psi \ra$  and the flux (in units of $\pi/2$)
through the triangle for $[{\pi\over2},{\pi\over2},0]$ (left three
columns) and $[{\pi\over2},-{\pi\over2},0]$ (right three columns) }
\centering
\begin{tabular}{|c | c c c | c c c| }
\hline $L$ & $\eta_{123}$ & $|\chi_{123}|$ & $|\Phi_{123}|$ &
$\eta_{123}$ & $|\chi_{123}|$ & $|\Phi_{123}|$ \\
\hline
2  & 0.376 & 0.356 & 0.483 & 0.368 & 0.325 & 0.461 \\
\hline
4  & 0.314 & 0.358 & 0.542 & 0.312 & 0.358 & 0.544 \\
\hline
\end{tabular}
\label{table:flux}
\end{table}

In the VMC calculation we take averages of $\la s_-|\psi \ra / \la
s|\psi \ra$ and $\la s_+|\psi \ra / \la s|\psi \ra$ separately. The
two quantities turn out to be complex conjugates with extremely high
accuracy, so one can denote the averaged $\la s_+ |\psi \ra / \la
s|\psi \ra$ and $\la s_- |\psi \ra / \la s|\psi \ra$ as $\la
\hat{\eta}_{123} \ra / 2 \pm i \la \hat{\chi}_{123} \ra/2$,
respectively. The flux $\Phi_{123}$ is deduced from $\tan (\Phi_{123}
) = \la \hat{\chi}_{123} \ra / \la \hat{\eta}_{123} \ra$. The signs
of the flux after the projection turned out to be in perfect accord
with the mean-field predictions. On the other hand, the amount of
flux is reduced from the mean-field value $\pi/2$ after the
projection (See Table \ref{table:flux}). The chirality was zero
within a statistical error for the non-chiral states, $[0,0,0]$ and
$[0,0,\pi]$. The increase of the average flux $\Phi$ with the system
size is consistent with the scenario of a long-range ordering of the
chirality in the ground state.

\begin{figure}[ht]
\begin{center}
\begin{tabular}{ccc}
\resizebox{27mm}{!}{\includegraphics{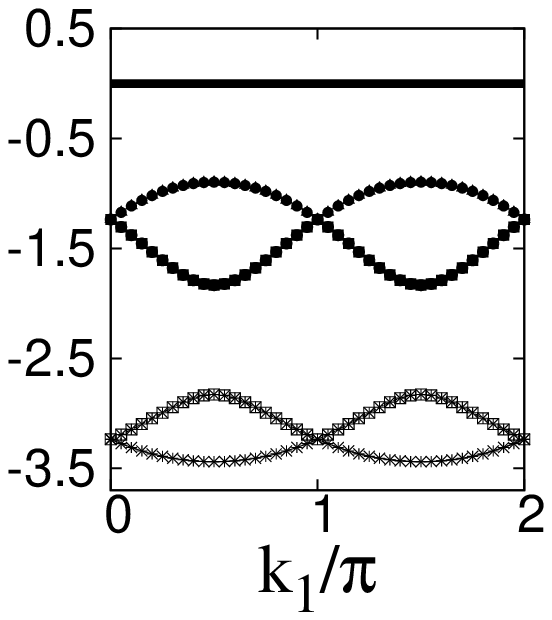}}
\resizebox{27mm}{!}{\includegraphics{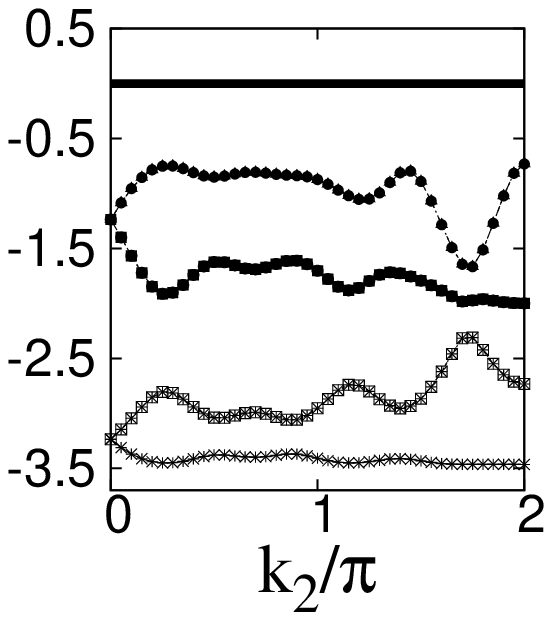}}
\resizebox{27mm}{!}{\includegraphics{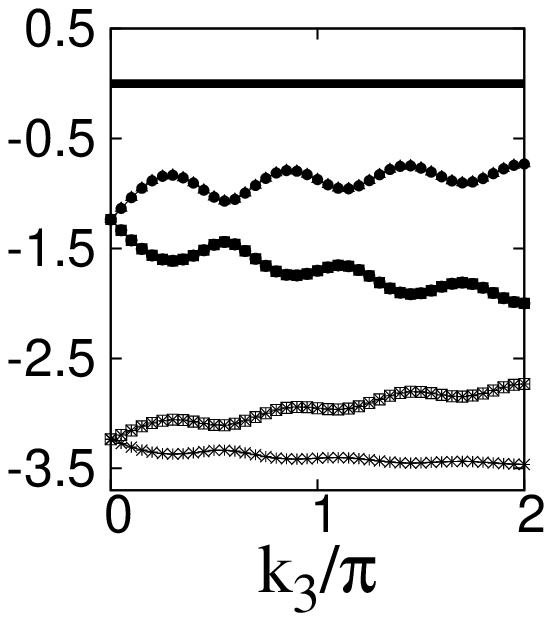}}
\end{tabular}
\end{center}
\caption{Mean-field energy bands for the $[{\pi\over2},-{\pi\over
2},0]$ state along $k_\alpha = \v k \cdot e_\alpha$ directions, $0
\le k_\alpha < 2\pi$. Each band is two-fold degenerate. The occupied
bands shown here are symmetric with the upper, unoccupied bands due
to particle-hole symmetry. The horizontal line $E_F =0$ is the Fermi
level. Dispersion along other directions (not shown) also show a full
gap.} \label{fig:chiral-bands}
\end{figure}

\textbf{Band Structure}: The band structures of the four flux
configurations in Fig.~\ref{fig:flux-configs} have been analyzed
along the three orthogonal directions $k_x, k_y, k_z$ as well as
along $k_\alpha$ defined to lie along the three $\hat{e}_\alpha$
directions, $\alpha = 1,2,3$. The relations between the two sets of
momenta are $k_1 \!=\! k_x$, $k_2 \!=\! k_x/2 \!+\! \sqrt{3}k_y/2$,
and $k_3 \!=\! k_x/2 \!+\! \sqrt{3}k_y/6 \!+\! \sqrt{6} k_z /3$. Flat
bands lying exactly at $E_F$ were observed in the $[0,0,0]$ and
$[{\pi\over2},{\pi\over2},0]$ states, and above $E_F$ for the
$[0,0,\pi]$ state. No flat bands exist for
$[{\pi\over2},-{\pi\over2},0]$. We describe the respective band
structures in more detail. (i) $[0,0,0]$: A two-fold degenerate flat
band lies exactly at $E_F$ for each $k_\alpha$ direction. The other
two non-degenerate bands lying below $E_F$ are dispersive. (ii)
$[{\pi\over 2},{\pi\over 2},0]$: A two-fold degenerate flat band at
$E_F =0$ was observed along the $k_z$ direction, $k_x = k_y = 0$. The
other two bands are given by $\sim \pm \cos (\sqrt{2/3}k_z )$. Along
each of the $k_\alpha$ directions, the uppermost and the lowermost
bands (which are related by particle-hole symmetry) are flat, and the
other two dispersive bands cross at $E_F$ as $k_\alpha$ equals a
multiple of $\pi$. There is no gap for this, or the $[0,0,0]$ flux
configuration. This explains the existence of a Fermi level
degeneracy in the finite-size mean-field energy spectra. (iii)
$[{\pi\over 2},-{\pi\over 2},0]$: Four doubly-degenerate bands lie
below the Fermi level and the others lie above it, separated by an
energy gap. Dispersion along the $k_\alpha$ directions are displayed
in Fig.~\ref{fig:chiral-bands}. (iv) $[0,0,\pi]$: Of the 16 bands,
the uppermost band lying above $E_F$ is 8-fold degenerate and flat
along each of the $k_\alpha$ directions. The remaining ones are the
four, 2-fold degenerate bands lying below the Fermi level, and
separated by a gap from the uppermost one. A gap separates the
occupied from the empty bands for the $[{\pi\over 2},-{\pi\over
2},0]$ and $[0,0,\pi]$ flux states, which also explains the absence
of a Fermi level degeneracy in an earlier mean-field calculation.

The presence of flat bands in the $[0,0,0]$ and $[{\pi\over
2},{\pi\over 2},0]$ flux states over a linear segment of the
Brillouin zone suggests that an instability might play a role. In the
case of the uniform flux state, the degeneracy-lifting terms are
given by the modulation of the bond amplitudes for triangles lying in
the plane spanned by $\hat{e}_1$ and $\hat{e}_2$ vectors.
Specifically it corresponds to $i \chi \rightarrow i(\chi \!+\!
\lambda)$ for the $(123)$ triangle of the up tetrahedron, and $i \chi
\rightarrow i(\chi \!-\!\lambda)$ for the (123) corners of the down
tetrahedron shown in Fig. \ref{fig:flux-configs} (a). The flux
through the triangles remain fixed at $\pi/2$. We observed that the
introduction of nonzero $\lambda$ did not decrease the variational
energy. Rather there was an increase in the third significant digit
of the energy value for $\lambda/\chi$ up to $0.1$ and by 2.7\% at
$\lambda/\chi=0.2$. The slow dependence of the overall energy is
partly due to the boundary conditions used, which already opened the
gap in the finite-size single-particle spectra. Another reason for
the lack of dependence may be that the degeneracy is confined to a
linear segment of the Brillouin zone, so that not enough states are
affected by the degeneracy-lifting mechanism. On the other hand, the
gapful nature of the band structures for $[{\pi\over 2},-{\pi\over
2},0]$ will guarantee stability of the Gutzwiller-projected state
against small fluctuations. At the moment we believe both types of
flux states to have a chance to represent the uniform ground states
of Eq. (\ref{eq:Heisenberg-H}).

\textbf{Discussion}: We find that the combined search using the
fermionic mean-field theory and Gutzwiller projection yields chiral
spin liquid states (both uniform and staggered flux types) with
ordered chiralities as the likely ground states of the $S=1/2$
Heisenberg spin Hamiltonian on the pyrochlore lattice. Previous
theories based on the perturbative expansion around the
single-tetrahedron
solution\cite{lacroix,harris,tsunetsugu,berg,moessner} did not find
such chiral spin states. Given the past claims of dimer instability,
a simultaneous search for a pronounced chirality and dimer
correlations in the exact diagonalization of Eq.
(\ref{eq:Heisenberg-H}) will be valuable in sorting out the
contending perspectives. The Fermi level degeneracy one finds in the
uniform flux state may have interesting consequences that will need
to be explored more carefully in the future.

\textit{Note added}: After the submission of this article, there
appeared a preprint\cite{sondhi} (arXiv:0809.0528v1) pertaining to
the chiral spin state on pyrochlore lattice. Their conclusions
overlap with ours.

\acknowledgments H.J.H. thanks Dung-Hai Lee and Ashvin Vishwanath for
discussions. This work was supported by the Korea Research Foundation
Grant (KRF-2008-521-C00085, KRF-2008-314-C00101) and in part by the
Asia Pacific Center for Theoretical Physics.

\end{document}